\documentclass[twocolumn,prd,10pt,nofootinbib,superscriptaddress]{revtex4-1}
\usepackage{}
\usepackage{epsfig,latexsym,amssymb}
\usepackage{latexsym}
\usepackage{amsmath}
\usepackage{amssymb}
\usepackage{amsfonts}
\usepackage{comment}
\usepackage{graphicx}
\usepackage{verbatim}
\usepackage{epstopdf}
\usepackage{hyperref, url}
\hypersetup{colorlinks   = true,
            urlcolor     = blue,
            citecolor    = blue,
            linkcolor    = blue,
            menucolor    = blue,
            anchorcolor  = blue,
            filecolor    = blue}

\newcommand{\be}{\begin{equation}}
\newcommand{\ee}{\end{equation}}
\newcommand{\bea}{\begin{eqnarray}}
\newcommand{\eea}{\end{eqnarray}}

\begin{document}

\title{Detection of dilute axion stars with stimulated decay}

\author{Haoran Di}
\email{hrdi@ecut.edu.cn}
\affiliation{School of Science, East China University of Technology, Nanchang 330013, China}
\author{Haihao Shi}
\email{shihaihao@xao.ac.cn}
\affiliation{Xinjiang Astronomical Observatory, CAS, Urumqi 830011, China}
\affiliation{College of Astronomy and Space Science, University of Chinese Academy of Sciences, Beijing 101408, China}
\author{Zhu Yi}
\email{yz@bnu.edu.cn}
\affiliation{Faculty of Arts and Sciences, Beijing Normal University, Zhuhai 519087, China}
\affiliation{Advanced Institute of Natural Sciences, Beijing Normal University, Zhuhai 519087, China}

\begin{abstract}
The anomalous orbits of trans-Neptunian objects can be accounted for by the planet 9 hypothesis. One intriguing possibility is that planet 9 could be a dilute axion star captured by the solar system, with the ratio of the axion star to dark matter being approximately 1/10. Although dilute axion stars can emit monochromatic signals through two-photon decay, the spontaneous decay signal is too weak to be detected by radio telescopes. However, we find that stimulated decay of the dilute axion star, which explains planet 9, can occur by directing a radio beam with a power of 50MW into the star. The resulting echo can be detected by terrestrial telescopes such as SKA, FAST, ngLOBO, and LOFAR. Therefore, the dilute axion star can be distinguished from other planet 9 candidates, such as a primordial black hole or a free-floating planet captured by the solar system.
\end{abstract}

\maketitle

\section{Introduction}
Accumulating evidence from diverse observations and theoretical models has led to the understanding that dark matter constitutes a significant fraction of the Universe's overall energy density. However, the precise nature and composition of dark matter particles remain enigmatic and elusive. The QCD axion \cite{Weinberg:1977ma,Wilczek:1977pj}, which emerges from the Peccei-Quinn mechanism \cite{Peccei:1977ur,Peccei:1977hh} that provides a prominent solution to the strong-CP problem, is a candidate for dark matter.
Additionally, within the framework of string theory, there are compelling reasons to believe in the existence of a diverse array of axionlike particles (ALPs) spanning a wide range of mass scales, a concept known as the ``axiverse" \cite{Arvanitaki:2009fg}. In this article, we collectively refer to QCD axions and ALPs as ``axions". Axions can be produced nonthermally
in the early Universe through various mechanisms, such as the misalignment mechanism \cite{Preskill:1982cy,Abbott:1982af,Dine:1982ah}, the decay of string defects \cite{Gorghetto:2020qws}, or the kinetic misalignment mechanism \cite{Co:2019jts}. Due to their bosonic nature, axions can achieve extremely high phase space densities, leading to the phenomenon of Bose-Einstein condensation (BEC) \cite{Sikivie:2009qn}. As a result of BEC, axions can form gravitationally bound structures known as axion stars.
In the post-inflationary scenario, more than ten percent of QCD axions can collapse into axion stars at matter-radiation equality \cite{Gorghetto:2024vnp}. For recent reviews of axion stars, see Refs. \cite{Braaten:2019knj,Visinelli:2021uve,Zhang:2024bjo}.

The observational anomalies related to the orbits of trans-Neptunian objects (TNOs) \cite{Brown:2004yy,Trujillo,Batygin:2016zsa} and a set of gravitational anomalies observed by the Optical Gravitational Lensing Experiment (OGLE) \cite{Mroz:2017mvf} require explanation. Primordial black holes (PBHs) \cite{Niikura:2019kqi}, or axion stars \cite{Sugiyama:2021xqg} with masses around $M\sim0.5$-$20M_\oplus$ can potentially explain the gravitational anomalies observed by OGLE. The planet 9 hypothesis, proposing a mass of 5-15$M_\oplus\approx$1.5-4.5$\times10^{-5}M_\odot$ at a distance of 300-1000$\rm{AU}$ from the Sun, was introduced to account for the anomalies in TNO orbits \cite{Batygin-PR}. An intriguing possibility is that planet 9 could be a PBH \cite{Scholtz:2019csj} or an axion star \cite{Di:2023xaw} captured by the solar system, with the proportion of dilute axion star to dark matter being $\Omega_{\rm{AS}}/\Omega_{\rm{DM}}\simeq 1/10$. Although axion stars can emit monochromatic signals through two-photon decay, these decay signals are too weak to be detected in the radio band.

Electromagnetic radiation with an angular frequency equal to half the axion mass has the capability to stimulate the decay of cold dark matter axions, leading to the production of an echo. This phenomenon presents a unique opportunity for the detection of axion dark matter. Therefore, axion dark matter can be searched for by emitting a powerful beam of microwave radiation into space and listening for its echo \cite{Arza:2019nta,Arza:2021nec,Arza:2022dng,Arza:2023rcs,Gong:2023ilg}. This method provides a promising approach to identifying and studying axion dark matter.

In this article, we propose that the stimulated decay \cite{Kephart:1994uy,Rosa:2017ury,Caputo:2018vmy,Carenza:2019vzg,Di:2023nnb} of a dilute axion star can occur by emitting a radio beam into the star, producing an echo detectable by terrestrial telescopes. This method allows for the differentiation of a dilute axion star from other planet 9 candidates, such as a PBH or a free-floating planet (FFP) captured by the solar system. Throughout this article, we use natural units where $c = \hbar=1$.

\section{Dilute Axion Stars}
The QCD axion is a pseudo-Nambu-Goldstone boson with spin 0, characterized by its lightness, weak self-interaction, and feeble interactions with Standard Model particles.
Its potential $V(\phi)$ must be periodic:
\bea
V(\phi)=V(\phi+2\pi f_a),
\eea
where $\phi$ represents the axions field, and $f_a$ is the axion decay constant, representing the energy scale at which the Peccei-Quinn symmetry $U(1)_{\rm{PQ}}$ spontaneously breaks. The common model for this potential is the instanton potential \cite{Peccei:1977ur}:
\bea
V(\phi)={m_\phi}^2 {f_a}^2\left[1-\cos\left(\frac{\phi}{f_a}\right)\right]
\eea
which can be expanded as
\bea
V(\phi)={1\over2} m_\phi^2 \phi^2+\frac{\lambda}{4!}\phi^4+...,
\eea
where $m_\phi$ is the axion mass, and $\lambda=-m_\phi^2/f_a^2$ represents the attractive self-interaction coupling constant.
Axions, being bosons that follow Bose-Einstein statistics, can achieve extremely high phase space density, resulting in the formation of BECs \cite{Sikivie:2009qn}. These BECs can form axion stars, which may exist in both dilute and dense configurations \cite{Chavanis:2017loo,Visinelli:2017ooc,Eby:2019ntd}. However, the lifespan of a dense axion star might be too short for it to have significant cosmological importance as an astrophysical entity \cite{Braaten:2019knj,Seidel:1991zh,Hertzberg:2010yz,Eby:2015hyx}.

A stable dilute axion star can be described as a balanced state where the attractive self-gravity of axion particles is counteracted by the repulsive gradient energy. This equilibrium can be sustained as long as the dilute axion star maintains a sufficiently low density, minimizing the impact of self-interactions. However, when the mass of the dilute axion star exceeds the critical value determined by the decay constant and mass of the axion, this equilibrium will be disrupted.
The maximum mass \cite{Chavanis:2011zi,Chavanis:2011zm} of a dilute axion star that can maintain stability, and the corresponding minimum radius, are given by
\bea
\label{maximum mass}
M_{\rm{max}}\sim 5.073\frac{M_{pl}}{\sqrt {|\lambda|}},~~~~~~R_{\rm{min}}\sim\sqrt {|\lambda|}\frac{M_{pl}}{m_\phi}\lambda_c,
\eea
where $M_{pl}$ represents the Planck mass and $\lambda_c$ denotes the axion's Compton wavelength.
When the dilute axion star's mass increases and surpasses the maximum mass given by Eq.\eqref{maximum mass} as a result of merger events \cite{Mundim:2010hi,Cotner:2016aaq,Schwabe:2016rze,Eby:2017xaw,Hertzberg:2020dbk,Du:2023jxh,Maseizik:2024qly} or the accretion of axions from the background \cite{Chen:2020cef,Chan:2022bkz,Dmitriev:2023ipv}, the self-interactions of axions become dominant and can cause the axion star to collapse \cite{Levkov:2016rkk,Eby:2016cnq,Fox:2023xgx}.
By substituting the attractive self-interaction coupling constant $\lambda=-m_\phi^2/f_a^2$ into Eq. \eqref{maximum mass}, we can obtain
\bea\label{critical mass}
M_{\rm{max}}\sim5.97\times10^{-6}M_{\odot}\left(\frac{m_\phi}{10^{-6}{\rm{eV}}}\right)^{-1}
\left(\frac{f_a}{10^{17}{\rm{GeV}}}\right),\nonumber\\
\eea
\bea\label{radius}
R_{\rm{min}}\sim2.41\times10^{-2}{\rm{km}}\left(\frac{m_\phi}{10^{-6}{\rm{eV}}}\right)^{-1}
\left(\frac{f_a}{10^{17}{\rm{GeV}}}\right)^{-1}.\nonumber\\
\eea
For general dilute axion stars, we denote the mass and radius as $M_{\rm{AS}}$ and $R_{\rm{AS}}$, respectively.

In the context of the instanton potential, the general Lagrangian of the axion can be expressed as follows:
\bea \label{axion}
{\cal L}&=&{1\over 2}\partial_{\mu}\phi\partial^{\mu}\phi-{1\over 2}m_{\phi}^2\phi^2
-\frac{\lambda}{4!}\phi^4 \\ \nonumber
&&+\frac{\alpha K}{8\pi f_a}\phi F_{\mu\nu} \tilde F^{\mu\nu}+...,
\eea
where $F_{\mu\nu}$ denotes the electromagnetic field strength tensor, $\tilde F^{\mu\nu}$ is the dual tensor of $F_{\mu\nu}$,  $\alpha$ is the fine structure constant, and $K$ is a model-dependent constant generically expected to be of order 1; we will set $K=1$ in this article. For QCD axions, their coupling
to gluons leads to a well-defined relationship between the decay constant $f_a$ and the mass of the axion \cite{Sikivie:2006ni}:
\bea \label{relation}
{m_{\phi}}\simeq6~{\rm{\mu eV}}\left(\frac{10^{12}{\rm{GeV}}}{f_a}\right),
\eea
as illustrated in Fig.~\ref{fig:fa}. In contrast, ALPs are not required to address the strong-CP problem and therefore do not interact directly with gluons. As a result, the mass $m_\phi$ and the decay constant $f_a$ are independent parameters for ALPs.
Axions are not entirely stable and can decay into two photons due to the interaction term ${\cal L}_{int}=\frac{\alpha K}{8\pi f_a}\phi F_{\mu\nu} \tilde F^{\mu\nu}$.  The interaction of axions with two photons underpins various astrophysical implications and guides experimental search strategies for detecting these elusive particles.
The decay rate of an axion, derived from the interaction term, is given by
\bea
\Gamma_\phi&=&\frac{\alpha^2 m_\phi^3}{256 \pi^3 f_a^2}\nonumber\\
&=&1.02\times 10^{-63} {\rm{s}^{-1}} \left( \frac{m_\phi}{10^{-6}\rm{eV}} \right)^3 \left(\frac{f_a}{10^{17}\rm{GeV}}\right)^{-2}.
\eea
The axion's lifetime is the inverse of its decay rate and is characterized by the following expression:
\bea
\tau_\phi=9.80\times 10^{62} {\rm{s}} \left( \frac {m_\phi} {10^{-6}\rm{eV}}\right)^{-3} \left(\frac{f_a} {10^{17}\rm{GeV}}\right)^2.
\eea
For axions with masses in the range of $10^{-8}{\rm{eV}}$ to $10^{-5}{\rm{eV}}$, which we are interested in, their lifetimes exceed the age of the universe. Therefore, the spontaneous decay of axion does not disrupt the stability of a dilute axion star in this mass range.

\section{Stimulated Decay from Dilute Axion Stars}
The Bose enhancement effect must be considered because the axion is a pseudoscalar boson with spin 0. We provide a brief overview of the stimulated decay of axions. The rate of change in the photon number density $n_\lambda$ over time within the dilute axion star, due to the process axion $\leftrightarrow\gamma\gamma$, is given by the Boltzmann equation \cite{Kephart:1994uy}:
\bea
\frac{dn_\lambda}{dt}&=&\int\frac{d^3p}{(2\pi)^32p^0} \int\frac{d^3k_1}{(2\pi)^32k_1^0} \int\frac{d^3k_2}{(2\pi)^32k_2^0}\nonumber\\
 &&\times(2\pi)^4\delta^4(p-k_1-k_2)|{\cal M}|^2\nonumber\\
&&\times\{f_\phi(\textbf{p})[1+f_\lambda(\textbf{k}_1)][1+f_\lambda(\textbf{k}_2)]\nonumber\\
&&-f_\lambda(\textbf{k}_1)f_\lambda(\textbf{k}_2)[1+f_\phi(\textbf{p})]\},
\eea
where $f_\phi$ and $f_\lambda$ represent the phase space densities of the axion and photon, respectively, and $\cal M$ is the matrix element determined by the interaction term ${\cal L}_{int}=\frac{\alpha K}{8\pi f_a}\phi F_{\mu\nu} \tilde F^{\mu\nu}$.
Assuming that the phase space distribution of axions and photons is approximately homogeneous and isotropic within the dilute axion star, to simplify the integration in the equation above, the expression for the variation of photon number density over time becomes \cite{Kephart:1994uy,Rosa:2017ury}
\bea\label{photon number density}
\frac{dn_\gamma}{dt}=\Gamma_\phi\left[2n_\phi
\left(1+\frac{8\pi^2}{m_\phi^3v}n_\gamma\right)
-\frac{16\pi^2}{3m_\phi^3}
\left(v+\frac{3}{2}\right)n_\gamma^2\right],\nonumber\\
\eea
where $n_\phi$ is the axion number density, and $v\sim1/(2R_{\rm{AS}}m_\phi)$ represents the maximum axion velocity within the dilute axion star, as determined by the Heisenberg uncertainty principle.

In Eq. \eqref{photon number density}, the initial terms within the square brackets represent spontaneous and stimulated decay. The final term, proportional to $n_\gamma^2$, indicates the process of inverse decay. The component within the final term, scaled by a factor of $3/2$, denotes the production of ``sterile" axions \cite{Kephart:1994uy}, which can escape from the dilute axion star.
Additionally, the photons generated by spontaneous and stimulated decay escape from the dilute axion star at a rate of
\bea \label{escape rate}
\Gamma_e=R_{\rm{AS}}^{-1},
\eea
which is essentially the inverse of the radius of the dilute axion star. Considering a dilute axion star of critical mass and substituting Eq. \eqref{radius} into Eq. \eqref{escape rate}, we obtain
\bea
\label{gamma}
\Gamma_e\sim1.24\times10^7{\rm{s^{-1}}}\left(\frac{m_\phi}{10^{-6}{\rm{eV}}}\right)
\left(\frac{f_a}{10^{17}{\rm{GeV}}}\right).
\eea
Therefore, by integrating Eq. \eqref{photon number density} and accounting for the escaped photons, we previously derived the following coupled differential equations that describe the changes in the number of photons and axions over time within the dilute axion star \cite{Di:2023nnb}:
\bea
\frac{dN_\gamma}{dt}=\Gamma_\phi[2N_\phi(1+AN_\gamma)-B N_\gamma^2]-\Gamma_e N_\gamma,
\eea
\bea
\frac{dN_\phi}{dt}=-\Gamma_\phi[N_\phi(1+A N_\gamma)-C N_\gamma^2],
\eea
where $A=6\pi(m_\phi^3vR_{\rm{AS}}^3)^{-1}$, $B=4\pi (v+3/2)(m_\phi^3R_{\rm{AS}}^3)^{-1}$, $C=4\pi v(m_\phi^3R_{\rm{AS}}^3)^{-1}$. From the above two equations, we can see that as long as the condition $AN_\gamma>1$ is met within the dilute axion star, stimulated decay dominates over spontaneous decay.
However, even for dilute axion stars with critical mass, the photons with a number $N_\gamma\simeq(2\Gamma_\phi/\Gamma_e)N_\phi$ present in the star arising from spontaneous decay cannot trigger the stimulated decay of axions because $N_\gamma A\ll1$. For example, $N_\gamma A\sim2.76\times10^{-6}$ with the parameter $m_\phi=10^{-6}{\rm{eV}}$ and $f_a=2.51\times10^{17}{\rm{GeV}}$.
\begin{figure}
\begin{center}
\includegraphics[width=0.45\textwidth]{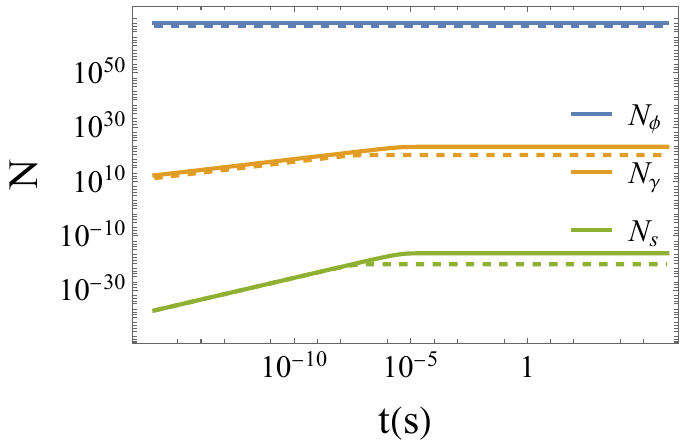}
\caption{Numerical evolution of the axion, photon, and ``sterile'' axion numbers for a dilute axion star with $M_{\rm{AS}}=1.5\times10^{-5}M_\odot$ under a powerful beam of microwave radiation of $50{\rm{MW}}$. The parameters corresponding to the solid line are $m_\phi=10^{-7}{\rm{eV}}$ and $f_a=2.51\times10^{16}{\rm{GeV}}$, while the parameters corresponding to the dashed line are $m_\phi=10^{-6}{\rm{eV}}$ and $f_a=2.51\times10^{17}{\rm{GeV}}$. Timing begins from the irradiation of radio waves on the dilute axion star. For the case of $m_\phi=10^{-6}{\rm{eV}}$, after approximately $10^{-7}$ seconds of irradiation, the stimulated decay of the dilute axion star reaches a stable state with a photon number $N_\gamma\sim10^{20}$.}
\label{fig:number}
\end{center}
\end{figure}

Now consider transmitting radio waves with a power of $P$ into the dilute axion star. The presence of a background of photons with energies near half the axion mass can greatly increase the axion decay rate. The coupled differential equations governing the evolution of the number of axions and photons within the dilute axion star then change to
\bea
\frac{dN_\gamma}{dt}=\Gamma_\phi\{2N_\phi[1+A(N_\gamma+N_{\gamma 0})]\nonumber\\
-B(N_\gamma+N_{\gamma 0})^2\}-\Gamma_e N_\gamma,
\eea
\bea
\frac{dN_\phi}{dt}=-\Gamma_\phi\{N_\phi[1+A(N_\gamma+N_{\gamma 0})]\nonumber\\-C(N_\gamma+N_{\gamma 0})^2\},
\eea
where $N_{\gamma 0}=2PR_{\rm{AS}}/m_\phi$ is the number of radio wave photons emitted from Earth within the dilute axion star.

Dilute axion stars tend toward a critical state through the accretion of axions from the background or through merger events. If they exceed this critical state, they will collapse, emitting strong radio signals \cite{Di:2023nnb} or relativistic axions \cite{Levkov:2016rkk}. After the radiation stops, a residual dilute axion star remains, with a mass that has not significantly decreased compared to the axion star in a critical state, placing it in a subcritical state. Therefore, the likelihood of dilute axion stars being in a critical or subcritical state is relatively high. For simplicity, in the following discussion, we will consider dilute axion stars to be in a critical state.
By substituting Eq. \eqref{radius} into $N_{\gamma 0}=2PR_{\rm{AS}}/m_\phi$, we obtain
\bea
N_{\gamma 0}&=&5.02\times10^{25}\left(\frac{P}{50\rm{MW}}\right)\nonumber\\
&&\times\left(\frac{m_\phi}{10^{-6}\rm{eV}}\right)^{-2}
\left(\frac{f_a}{10^{17}\rm{GeV}}\right)^{-1}.
\eea
As long as the condition $A(N_\gamma+N_{\gamma 0})>1$ is met, stimulated decay dominates over spontaneous decay. Considering the transmission of  radio waves with a power of $P={50\rm{MW}}$ into the dilute axion star, we get $N_{\gamma 0}=2.00\times10^{25}$ and $AN_{\gamma 0}=3.18\times10^{23}\gg1$ with the parameters $m_\phi=10^{-6}{\rm{eV}}$ and $f_a=2.51\times10^{17}{\rm{GeV}}$.  The high photon occupation number of the beam results in a Bose enhancement effect on the axion's decay to two photons. Therefore, using radio waves of appropriate power to irradiate a dilute axion star can indeed induce stimulated decay.

\begin{figure}
\begin{center}
\includegraphics[width=0.45\textwidth]{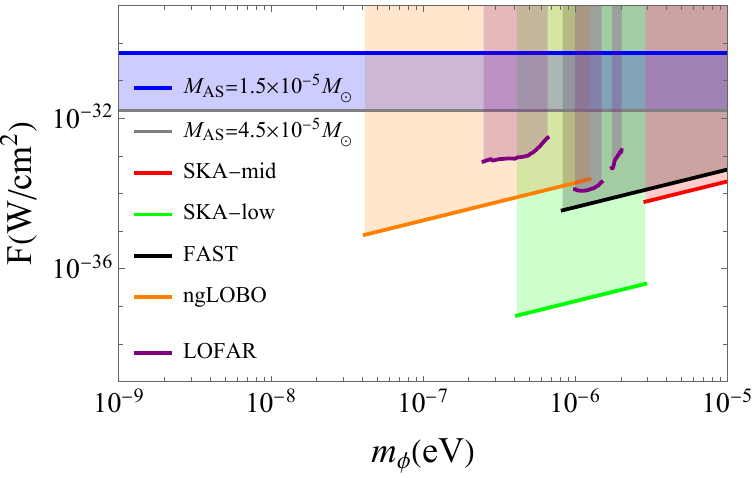}
\caption{Flux of the dilute axion star compared to the minimum detectable flux of SKA, FAST, ngLOBO, and LOFAR with an observation time of 1 hour. Due to the uncertainty in the mass of planet 9, estimated to be between 5 and 15 Earth masses (5-15$M_\oplus\approx $1.5-4.5$\times10^{-5}M_\odot$), the flux of the dilute axion star is represented within the range of blue-shaded areas. The uncertainty in distance has also been taken into account, as smaller mass axion stars correspond to a closer distance of 300 AU, while larger mass axion stars are associated with a distance of 1000 AU. This ensures that the gravitational influence of the axion star on TNOs remains consistent.}
\label{fig:flux}
\end{center}
\end{figure}

\begin{figure}
\begin{center}
\includegraphics[width=0.45\textwidth]{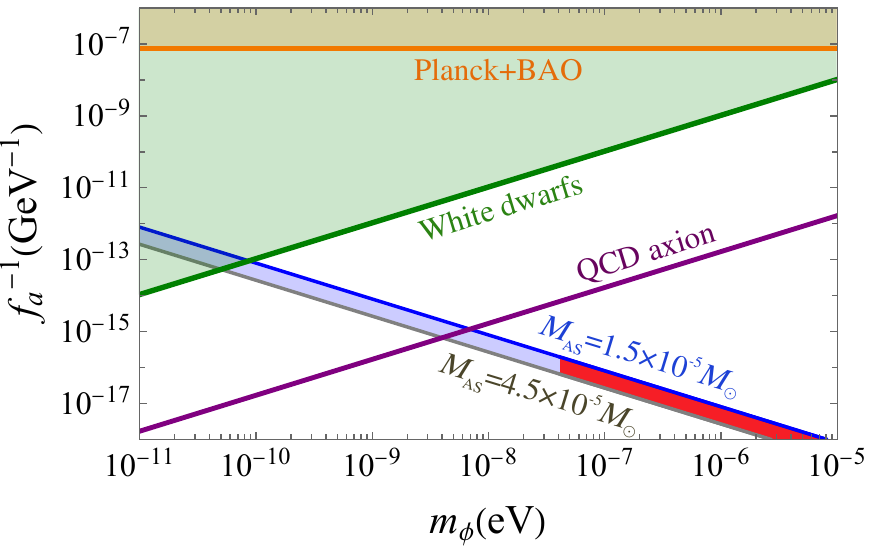}
\caption{The parameter space of axions.  The parameter space above the orange solid line is excluded by constraints from Planck 2018 combined with BAO measurements \cite{Caloni:2022uya}. Exclusion from white dwarf observations \cite{Balkin:2022qer} are indicated by the green area. The purple line represents QCD axions, which are constrained by Eq. \eqref{relation}. The red shaded area represents the parameter space that can be jointly explored by radio telescopes such as SKA, FAST, LOFAR, and ngLOBO. If a non-critical axion star is considered to explain planet 9, then the parameter space below the red region remains viable, and some of this space may also be scanned by radio telescopes.}
\label{fig:fa}
\end{center}
\end{figure}

In Fig.~\ref{fig:number}, we present a numerical solution for two representative sets of parameters: the solid line corresponds to $m_\phi=10^{-7}{\rm{eV}}$ and the decay constant $f_a=2.51\times10^{16}{\rm{GeV}}$, and the dashed line corresponds to $m_\phi=10^{-6}{\rm{eV}}$ and $f_a=2.51\times10^{17}{\rm{GeV}}$.
The luminosity of the dilute axion star is given by
\bea
L_\phi=\frac{1}{2}\Gamma_e m_\phi N_\gamma.
\eea
An interesting feature of stimulated decay is that the emitted photons are confined to travel either along or in the opposite direction to the incident photons in the axion rest frame. Nonetheless, the actual velocity distribution of axions within the dilute axion star causes the emitted photons to spread out in space.
Therefore, the luminosity flux of the dilute axion star as perceived from Earth is approximatively given by the formula:
\bea
\label{flux}
F_\phi\sim{L_\phi\over 4\pi r_9^2}=\frac{\Gamma_e m_\phi N_\gamma}{8\pi r_9^2},
\eea
where $r_9\sim 300$-$1000\rm{AU}$ represents the distance of the dilute axion star to Earth. Substituting Eq. \eqref{gamma} into Eq. \eqref{flux} yields
\bea
\label{flux2}
F_\phi&\sim&3.89\times10^{-31}\left(\frac{m_\phi}{10^{-6}{\rm{eV}}}\right)^2\left(\frac{f_a}{10^{17}{\rm{GeV}}}\right)\nonumber\\
&&\times\left(\frac{N_\gamma}{10^{20}}\right)\left(\frac{300\rm{AU}}{r_9}\right)^2{\rm{W/cm^2}}.
\eea
Considering that planet 9 is composed of a dilute axion star with critical mass, we have the following equation:
\bea\label{M9}
M_9\sim5.97\times10^{-6}M_{\odot}\left(\frac{m_\phi}{10^{-6}{\rm{eV}}}\right)^{-1}
\left(\frac{f_a}{10^{17}{\rm{GeV}}}\right).\nonumber\\
\eea
The mass of planet 9, estimated to be between 5 and 15 Earth masses (5-15$M_\oplus\approx$1.5-4.5$\times10^{-5}M_\odot$), combined with Eq. \eqref{M9}, defines the parameter space constraints for axions, as illustrated in Fig.~\ref{fig:fa}. Substituting Eq. \eqref{M9} into Eq. \eqref{flux2} with $M_9\approx1.5\times10^{-5}M_\odot$ yields
\bea
\label{flux3}
F_\phi&\sim&9.78\times10^{-31}\left(\frac{m_\phi}{10^{-6}{\rm{eV}}}\right)^3\nonumber\\
&&\times\left(\frac{N_\gamma}{10^{20}}\right)\left(\frac{300\rm{AU}}{r_9}\right)^2{\rm{W/cm^2}},
\eea
corresponding to the blue soild line as illustrated in Fig.~\ref{fig:flux}. The spectral line of decay signal is nearly monochromatic with a frequency $f\simeq m_\phi/({4\pi})\approx1.21\times10^2(m_\phi/10^{-6}{\rm{eV}}){\rm{MHz}}$, which is a distinct characteristic signal.

The spectral flux density $S_{\rm{min}}$ is a critical parameter in radio astronomy, defining the minimum detectable signal strength for a radio telescope observing standard electromagnetic signals. It is calculated using the equation
\bea
S_{\rm{min}}&=&\frac{\rm{SEFD}}{\eta_s \sqrt{n_{\rm{pol}}\Delta\nu t_{\rm{obs}}}},
\eea
where SEFD represents the system's equivalent flux density, $\eta_s$ is the system
efficiency, $n_{\rm{pol}}$ is the number of polarizations, $\Delta\nu$ denotes the bandwidth, and $t_{\rm{obs}}$ represents the observation time.

The Five-hundred-meter Aperture Spherical radio Telescope (FAST) is the world's largest filled-aperture radio telescope, operating across a frequency range of 0.1-3$\rm{GHz}$ with a bandwidth of 800MHz, an SEFD of $2.2\rm{Jy}$, and $n_{\rm{pol}}=38$ \cite{SKA:2015}. FAST achieves a spectral flux density $S_{\rm{min}}=3.60\times10^{-33}\rm{W/m^2/Hz}$ with $\eta_s\simeq0.6$ and $t_{\rm{obs}}=1 \rm{hour}.$ In the context of the dilute axion star, assuming nonrelativistic axions with minimal velocity dispersion $\delta v\ll 1$ results in a narrow spectral line spread. Given a natural width of $10^{-6}f$ for the spectral line of the resulting echo, the minimum detectable flux $F_{\rm{min}}$ can be expressed as
\bea
F_{\rm{min}}&=&3.60\times10^{-43}\left(\frac{f}{\rm{Hz}}\right)\rm{W/cm^2}\nonumber\\
&=&4.36\times10^{-35}\left(\frac{m_\phi}{10^{-6}\rm{eV}}\right)\rm{W/cm^2},
\eea
illustrating FAST's sensitivity. Compared to Eq. \eqref{flux3}, $F_{\rm{min}}$ is significantly smaller than $F_{\phi}$, as shown in Fig.~\ref{fig:flux}, enabling FAST to detect the decay signal.

For a radio telescope array, the RMS sensitivity of flux density is given by
\bea
\Delta S_\nu=\frac{\rm{SEFD}}{\sqrt{n(n-1)t_{\rm{obs}}\Delta\nu}},
\eea
where ${\rm{SEFD}}={2kT_{sys}}/A_e$, $k$ is the Boltzmann constant, $T_{sys}$ is the system noise, $A_e$ is the effective area, and $n$ is the number of telescopes in the array.
The Square Kilometer Array (SKA) includes SKA1-Low and SKA1-Mid telescopes.  SKA1-Low covers the frequency of 0.05-0.35$\rm{GHz}$, with $\Delta\nu=300 \rm{MHz}$, $n=131072$, $\rm{SEFD}=4.9\rm{Jy}$ and $A_e/T_{sys}=559{\rm{m^2/K}}$. SKA1-Mid covers the frequency of 0.35-14$\rm{GHz}$, with $\Delta\nu=770 \rm{MHz}$, $n=197$, $\rm{SEFD}=1.8\rm{Jy}$ and $A_e/T_{sys}=1560{\rm{m^2/K}}$.
See Ref. \cite{SKA:2015} for additional details on SKA parameters.
Thus, we obtain $\Delta S_\nu=1.14\times10^{-35}\rm{W/m^2/Hz}$ for SKA1-Low and $\Delta S_\nu=1.74\times10^{-33}\rm{W/m^2/Hz}$ for SKA1-Mid with an observation time $t_{\rm{obs}}=1 \rm{hour}$. Therefore, the minimum detectable flux for SKA1-Low and SKA1-Mid is
$F_{\rm{min}}=1.38\times10^{-37}\left({m_\phi}/{10^{-6}\rm{eV}}\right)\rm{W/cm^2}$
and
$F_{\rm{min}}=2.11\times10^{-35}\left({m_\phi}/{10^{-6}\rm{eV}}\right)\rm{W/cm^2}$,
respectively, as shown in Fig.~\ref{fig:flux}.

The low frequency range of the next-generation LOw Band Observatory (ngLOBO) is 5-150$\rm{MHz}$, and the RMS sensitivity of flux density for one beam of ngLOBO with $t_{\rm{obs}}=1 \rm{hour}$ is $0.64\rm{mJy}$ \cite{ngLOBO:taylor2017}. Therefore, we can determine the RMS sensitivity of flux density as
$\Delta S_\nu=1.6\mu{\rm{Jy}}=1.6\times10^{-32}\rm{W/m^2/Hz}$. Thus, the minimum detectable flux of ngLOBO is
$F_{\rm{min}}=1.94\times10^{-34}\left({m_\phi}/{10^{-6}\rm{eV}}\right)\rm{W/cm^2}$,
as shown in Fig.~\ref{fig:flux}.
The Low Frequency ARray (LOFAR) is located in Europe, with core stations and the correlator situated in the Netherlands. LOFAR can detect radio signals in the frequency ranges of 30-80$\rm{MHz}$, 120-180$\rm{MHz}$, and 210-250$\rm{MHz}$. The SEFD of LOFAR as a function of frequency for the various operating bands and station configurations is provided in \cite{LOFAR:2013jil}. Using this information, we can determine the minimum detectable flux of LOFAR for an observation time of 1 hour, as shown in Fig.~\ref{fig:flux}.

Therefore, the stimulated decay of a dilute axion star can be triggered by emitting a radio beam with a power of 50MW into the star. The decay signal can be detected by radio telescopes such as SKA, FAST, ngLOBO, and LOFAR, as illustrated in Fig.~\ref{fig:flux}. The parameter space for axions that can be explored is shown in Fig.~\ref{fig:fa}. Consequently, a dilute axion star can be distinguished from other candidates for planet 9, such as a PBH or a FFP captured by the solar system.

The results above assume that axion stars can receive radio signals with a power of 50 MW at a maximum distance of 1000 AU. Currently, the multi-beam relativistic klystron amplifier can generate radio waves with an output power of 1.047 GW \cite{sun:2023}, much higher than the required 50 MW. This high power output makes it a promising device for emitting radio waves to irradiate axion stars. The radio waves produced by this device can be directed into a parabolic antenna to form a highly directional radio beam.
The power received by an axion star could be further enhanced by increasing the number of transmitters or by narrowing the divergence angle of the emitted radio waves. For an array with N transmitting devices, the power received by the dilute axion star is
\bea
P= \frac{N P_t G_t}{4\pi r_9^2} \pi R_{\rm{AS}}^2= \frac{N P_t G_t R_{\rm{AS}}^2}{4 r_9^2},
\eea
where  $P_t$ is the power input to the transmitting antenna, and $G_t$ is the antenna gain. The gain $G_t$ is related to the divergence angle of the emitted radio waves; smaller divergence angles result in greater gains.
For an array with $N=1000$, $P_t=1$ GW, and $G_t=220$ dBi, the power received by an axion star 1000 AU away is approximately 65 MW, given parameters $m_\phi=10^{-6}{\rm eV}$ and $f_a=10^{17}{\rm GeV}$, satisfying the necessary requirements.
Therefore, constructing an array of 1000 transmitters, each with a power output of 1 GW and an antenna gain of 220 dBi, could allow a potential axion star at 1000 AU to receive radio waves with a power around 50 MW. However, achieving an extremely high gain of 220 dBi experimentally requires further investigation.

\section{Conclusions}
The QCD axion or ALP is a prominent candidate for dark matter. Axions can collectively form a bound state known as an axion star through BEC. It is possible that a significant fraction of dark matter is composed of these axion stars. The anomalous orbits of TNOs can be explained by the planet 9 hypothesis. An intriguing possibility is that planet 9 could be a dilute axion star captured by the solar system, with the proportion of dilute axion stars to dark matter being $\Omega_{\rm{AS}}/\Omega_{\rm{DM}}\simeq 1/10$. Although axion stars can emit monochromatic signals through two-photon decay, the spontaneous decay signal is too weak to be detected. However, we find that the stimulated decay of a dilute axion star can be triggered by emitting a radio beam with a power of 50MW into the star. The resulting echo can be detected by terrestrial telescopes such as SKA, FAST, ngLOBO, and LOFAR. This echo from stimulated decay would provide a clear and distinctive signal, differentiating it from other forms of background noise and radiation. Therefore, a dilute axion star can be distinguished from other planet 9 candidates, such as a PBH or a FFP captured by the solar system. The exploration of axion stars offers a fascinating intersection of quantum field theory and gravitational physics, potentially unlocking new understanding of the Universe's dark sector.

\section{Acknowledgments}
H. Di would like to thank Shaowei Jia, Qin Fei and Ning Dai for useful discussions. This work was supported by National Natural Science Foundation of China under Grant No. 11947031 and East China University of Technology Research Foundation for Advanced Talents under Grant No. DHBK2019206.

\end{document}